\documentclass[aps,prb,twocolumn,superscriptaddress]{revtex4-1}
\usepackage{graphicx}
\usepackage{latexsym}
\usepackage{amsmath}
\usepackage{txfonts}
\usepackage{helvet}
\usepackage{braket}
\usepackage{amscd}

\usepackage{longtable}

\begin{document}

\title{Two-step gap opening across the quantum critical point in a Kitaev honeycomb magnet}

\author{Yuya Nagai}
\affiliation{Department of Physics, Graduate School of Science, Nagoya University, Furo-cho, Chikusa-ku, Nagoya 464-8602, Japan.}
\author{Takaaki Jin-no}
\affiliation{Technical Center, Nagoya University, Furo-cho, Chikusa-ku, Nagoya 464-8601, Japan.}
\author{Junki Yoshitake} 
\affiliation{Department of Applied Physics, University of Tokyo, Hongo, 7-3-1, Bunkyo, Tokyo 113-8656, Japan.}
\author{Joji Nasu} 
\affiliation{Department of Physics, Tokyo Institute of Technology, Ookayama, 2-12-1, Meguro, Tokyo 152-8551, Japan.}
\author{Yukitoshi Motome} 
\affiliation{Department of Applied Physics, University of Tokyo, Hongo, 7-3-1, Bunkyo, Tokyo 113-8656, Japan.}
\author{Masayuki Itoh}
\affiliation{Department of Physics, Graduate School of Science, Nagoya University, Furo-cho, Chikusa-ku, Nagoya 464-8602, Japan.}
\author{Yasuhiro Shimizu}
\affiliation{Department of Physics, Graduate School of Science, Nagoya University, Furo-cho, Chikusa-ku, Nagoya 464-8602, Japan.}

\date{\today}

\begin{abstract} 
Quantum spin liquid involves fractionalized quasipariticles such as spinons and visons. They are expressed as itinerant Majorana fermions and $Z_2$ fluxes in the Kitaev model with bond-dependent exchange interactions on a honeycomb spin lattice\cite{Kitaev}. The observation has recently attracted attention for a candidate material $\alpha$-RuCl$_3$, showing spin liquid behaviour induced by a magnetic field\cite{Plumb, Kubota, Sandilands, Chaloupka, Banerjee, Banerjee2, Do, Wolter, Baek, Sears, Ponomaryov, Winter}. Since the observable spin excitation is inherently composed of the two quasiparticles, which further admix each other by setting in the magnetic field as well as non-Kitaev interactions\cite{Knolle, Song, Yoshitake, Gohlke}, their individual identification remains challenging. Here we report an emergent low-lying spin excitation through nuclear magnetic and quadrupole resonance measurements down to $\sim 0.4$ K corresponding to $1/500$ of the exchange energy under the finely tuned magnetic field across the quantum critical point. We determined the critical behaviour of low-lying excitations and found evolution of two kinds of the spin gap at high fields. The two excitations exhibit repulsive magnetic field dependence, suggesting anti-crossing due to the hybridization between fractionalized quasiparticles. 
\end{abstract}


\maketitle


	Quantum liquids involve quasiparticles following fractional statistics, as known in the fractional quantum Hall effect for charge degrees of freedom\cite{Laughlin}. On Mott insulators, the spin excitation of quantum spin liquids without long-range ordering can be fractionalized into spinons and visons corresponding to itinerant Majorana fermions (MF) and localized $Z_2$ gauge fluxes for the soluble Kitaev model $\mathcal H = \Sigma_{(i,j)}\Sigma_\gamma J_K S^\gamma_i S^\gamma_j$, where $J_K$ is the bond-dependent exchange coupling between two spins along directions $\gamma (= x, y, z)$ on the honeycomb lattice (Fig. 1{\bf a})\cite{Kitaev, Jackeli}. Although MF and $Z_2$ fluxes do not carry electric charges, they can display unique thermodynamics and spin dynamics. For example, they are individually excited at characteristic temperatures, $T_{\rm H} \simeq 0.375J_K$ and $T_{\rm L} \simeq 0.012J_K$, leading to a two-step entropy release\cite{Nasu}. The MF excitation gives a gapless broad continuum up to the energy scale of $J_{\rm K}$ in the excitation spectrum. Despite a flux gap below $T_{\rm L} \simeq 0.012J_K$ (Ref. \onlinecite{Knolle, Yoshitake, Yoshitake2, Yoshitake3}), the energy scale is usually too small to be identified in real materials because of long-range ordering at low temperatures. 

A honeycomb-lattice magnet $\alpha$-RuCl$_3$ attracts salient interests as an example of the Kitaev model\cite{Plumb, Sandilands, Kubota}. The step-wise entropy release\cite{Kubota, Nasu, Do} and the excitation continuum\cite{Banerjee, Sandilands, Do, Yoshitake} above the anitiferromagnetic ordering temperature $T_{\rm N} = 7$--14 K highlight the Kitaev physics. The spin liquid ground state appears as the in-plane magnetic field exceeds a critical value $H_c = 7$--$8$ T (Ref. \onlinecite{Baek, Wolter, Sears, Winter, Wang}), above which the spin gap grows with increasing magnetic field\cite{Ponomaryov, Baek, Wolter, Sears, Wang}. However, the field dependence of the spin excitation remains controversial; for experiments including the electron spin resonance spectroscopy\cite{Ponomaryov} and the specific heat\cite{Sears,Wolter}, the spin gap persists down to zero field or vanishes around $H_c$. Theoretically, the MF excitation becomes also gapped under the magnetic field\cite{Kitaev} and constructs a non-trivial admixture with the flux excitation\cite{Winter, Song}. 

We investigate low-energy spin excitations via nuclear magnetic resonance (NMR) and nuclear quadrupole resonance (NQR) on $\alpha$--RuCl$_3$, which are complementary to high energy probes such as inelastic neutron scattering measurements. The NMR Knight shift and nuclear spin-lattice relaxation rate $T_1^{-1}$ measure the static spin susceptibility $\chi$ and the dynamical one $\chi({\bf q}, \omega)$ with the wave vectors {\bf q} and the NMR frequency $\omega$, respectively, which characterize fractional excitations of the Kitaev spin liquid at finite temperatures\cite{Yoshitake} and an effect of non-Kitaev interactions\cite{Song, Winter}. Previous NMR measurements conducted at high fields ($H > H_c$) and high temperatures ($T > 1.5$ K) uncovered the significant field dependence of $T_1^{-1}$, highlighting a possible spin liquid state. However, the detailed field evolution of quasiparticles is still in debate: e.g., the spin gap increases linearly above $H_c$ (Ref. \onlinecite{Baek}) or with a cubic law across $H_c$ (Ref. \onlinecite{Jansa}), or is absent\cite{Zheng}. For solving the discrepancy, we conducted comprehensive measurements including zero-field NQR and finely field-tuned NMR down to low temperatures ($ > 0.4$ K) for observing the quasiparticles from the ground state.

	\begin{figure}
	\begin{center}
	\includegraphics[scale=0.55]{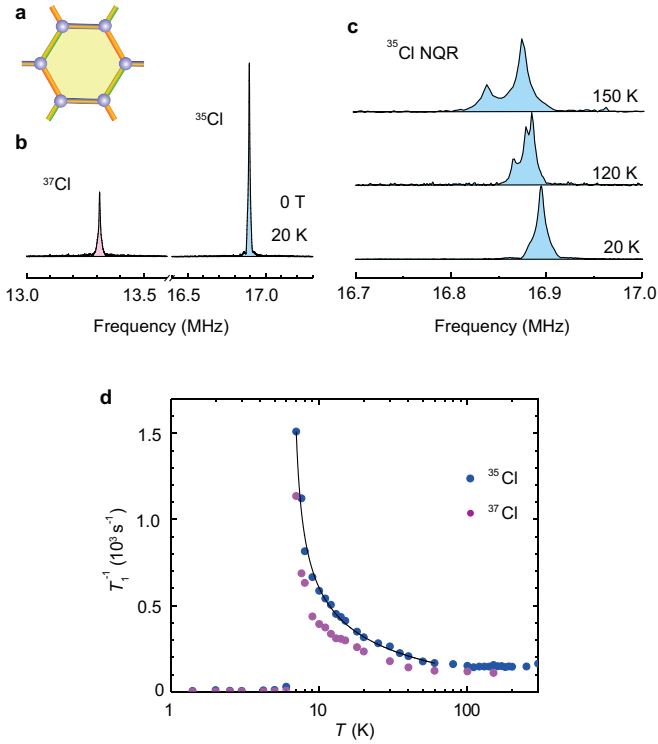}
	\end{center}
	\caption{\label{Fig1} 
	{\bf Zero-field NQR on a honeycomb magnet $\alpha$--RuCl$_3$}. {\bf a}, Kitaev model with bond-dependent exchange interactions on the honeycomb lattice. {\bf b}, $^{35}$Cl and $^{37}$Cl NQR spectra at 20 K. {\bf c}, Temperature $T$ dependence of $^{35}$Cl NQR spectrum for cooling process across the structural phase transition around 120 K. {\bf d}, Nuclear spin-lattice relaxation rate, $T_1^{-1}$, measured for $^{35}$Cl and $^{37}$Cl nuclei, which well scales to the square of the nuclear gyromagnetic ratio, $\gamma_n^2$ (see also Fig. S1). Solid curve: a fitting result by a scaling function $T_1^{-1} \sim (T/T_{\rm N} - 1)^{z^\prime}$ with $T_{\rm N} = 6.5 \pm 0.2$ K and the critical exponent $z^\prime = 0.45 \pm 0.03$. 
	}
	\end{figure}	

First we show the result of $^{35}$Cl and $^{37}$Cl NQR in the absence of magnetic field (Fig. 1). A single NQR spectrum (Fig. 1{\bf b}) for each nucleus ($I = 3/2$) with extremely sharp linewidth ($\sim$9 kHz) means a single Cl site in the crystal structure as expected for a rhombohedral $R\bar{3}$ lattice\cite{Morosin}. The resonance frequency (16.89 MHz, 13.31 MHz) and the spectral intensity well scale to the electric quadrupole moment ($^{35}Q = -8.2 \times 10^{-26}$, $^{37}Q = -6.5 \times 10^{-26}$ cm$^{2}$) and the natural abundance ($^{35}W = 75\%$, $^{37}W = 25\%$), respectively. Above 150 K, the spectrum splits into two with the intensity ratio of 2:1 (Fig. 1{\bf c}), consistent with two Cl sites in the monoclinic $P2_1/c$ structure\cite{Morosin}. The two phases coexist on the first-order structural transition around 120 K. 

At zero field, $T_1^{-1}$ measured for $^{35}$Cl and $^{37}$Cl (Fig. 1{\bf d}) is independent of temperature $T$ above 100 K, where the values for two nuclei scale to the square of the nuclear gyromagnetic ratio $\gamma_{\rm n}$ due to dominant magnetic fluctuations (Fig. S1). The similar behaviour was observed for $^{35}$Cl NMR (Fig. 2). The constant value ($T_1^{-1}$ = 107 s$^{-1}$ for $H \|c$) yields an effective exchange coupling $J = 170$ K (Ref. \onlinecite{Moriya}, Supplementary Information), in agreement with the previous estimate\cite{Do, Jansa}. With decreasing $T$, $T_1^{-1}$ starts to increase below 60 K. Instead of the broad peak of $T_1^{-1}$ as expected in the Kitaev model\cite{Yoshitake}, we observed a divergent enhancement with a scaling form $T_1^{-1}\sim (T/T_{\rm N}-1)^{z^\prime}$ toward $T_{\rm N}$ = 6.5 K. The critical exponent $z^\prime = 0.45 \pm 0.02$ is distinct from that of the two-dimensional Ising antiferromagnet ($z^\prime = 0.75$)\cite{Hohenberg,Dutta}. Therefore, $T_1^{-1}$ just above $T_{\rm N}$ is likely governed by spin fluctuations at a finite wave vector ($q \neq 0$) toward magnetic ordering. 

	\begin{figure}
	\begin{center}
	\includegraphics[scale=0.6]{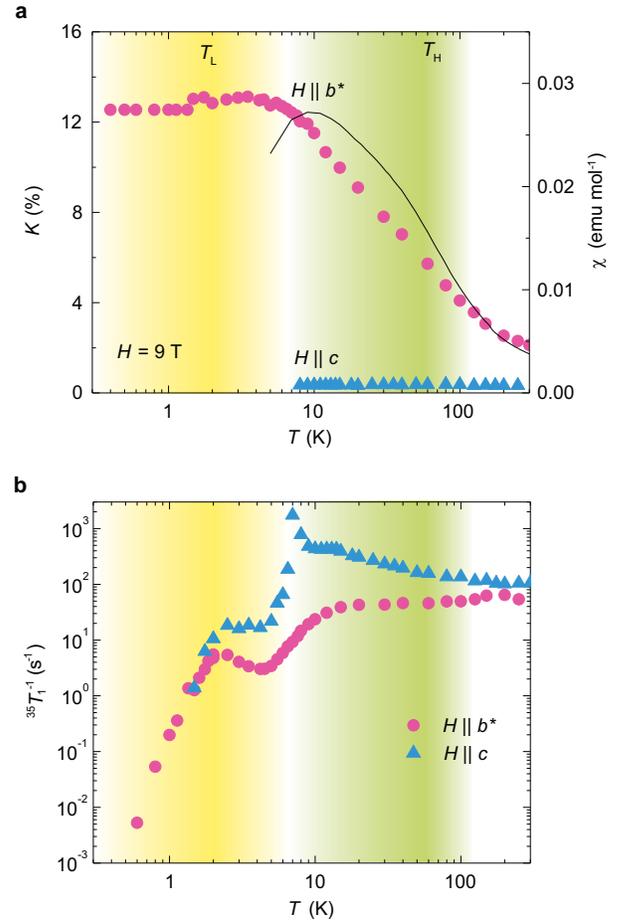}
	\end{center}
	\caption{\label{Fig2} 
	{\bf Anisotropic static and dynamical spin susceptibility}. {\bf a}, Temperature dependence of Knight shifts $K_c$ and $K_{b^*}$ at 9 T applied parallel and perpendicular to the $c$ axis, in comparison with the magnetic susceptibility $\chi$ (solid curve, the right-hand axis) at 1.0 T applied along the $ab$ plane. {\bf b}, Nuclear spin-lattice relaxation rate $T_1^{-1}$ for $H \parallel c$ and $b^*$, which measures the dynamical spin susceptibility normal to the field, $\chi_{\perp}({\bf q}, \omega)$. $T_{\rm H}$ and $T_{\rm L}$ denote the temperatures where MF and $Z_2$ fluxes are thermally excited, respectively\cite{Yoshitake}.  
	}
	\end{figure}

	Under the magnetic field of 9.0 T, we measured the $^{35}$Cl Knight shift, $K_c$ and $K_{b^*}$, for the out-of-plane $c$ axis and the in-plane $b^*$ axis normal to the $a$ axis, which displays highly anisotropic behaviour (Fig. 2{\bf a}). Here we evaluated the Knight shift by simulating the NMR spectrum using the nuclear quadrupole frequency $\nu_{\rm Q}$ obtained from the NQR measurement (Figs. S2--S4). We find that $K_c$ is independent of $T$, while $K_{b^*}$ is largely enhanced at low temperatures ($T \lesssim T_{\rm H} = 60$ K), as observed in the bulk magnetic susceptibility $\chi$, and then it levels off below $T^* \sim$ 10 K $\simeq  0.06J$. Instead of the magnetic ordering, $K_{b^*}$ shows a broad maximum around 4 K and becomes $T$-independent for $T \lesssim T_{\rm L} = 2$ K without an indication of magnetic ordering down to 0.4 K ($\simeq 0.002J_K$). In comparison with the numerical calculation\cite{Yoshitake, Yoshitake2, Yoshitake3}, the constant $K_{b^*} = 12.5\%$ corresponding to 0.028 emu mol$^{-1}$ is located in between antiferromagnetic ($0.0058$ emu mol$^{-1}$) and ferromagnetic ($0.093$ emu mol$^{-1}$) cases, indicating the significant ferromagnetic spin correlation. 
	
The $T$ dependence of $T_1^{-1}$ is also highly anisotropic, as shown in Fig. 2{\bf b}. In general, $T_1^{-1}$ monitors spin fluctuations with wave vectors ${\bf q}$ normal to the external field. In the present case, owing to the strong easy-plane anisotropy, in-plane spin fluctuations dominate $T_1^{-1}$ for both $H \parallel  c$ and $b^*$. and hence the observed anisotropy of $T_1^{-1}$ should come from the dependence of spin excitations against the in-plane field component. Indeed, for $H \parallel  c$ without the in-plane component, $T_1^{-1}$ exhibits similar behaviour to that of zero field, showing a sharp peak at $T_{\rm N}$ = 6.5 K. For $H \parallel b^*$, $T_1^{-1}$ is nearly independent of $T$ down to 20 K and then becomes suppressed, pointing to a gap opening over the ${\bf q}$ space. Upon further cooling, a prominent peak appears around $T_{\rm L}$, followed by a steep suppression. The result signifies that there are two kinds of the spin gap in the excitation spectrum of the field-driven spin liquid state.

	\begin{figure}
	\begin{center}
	\includegraphics[scale=0.6]{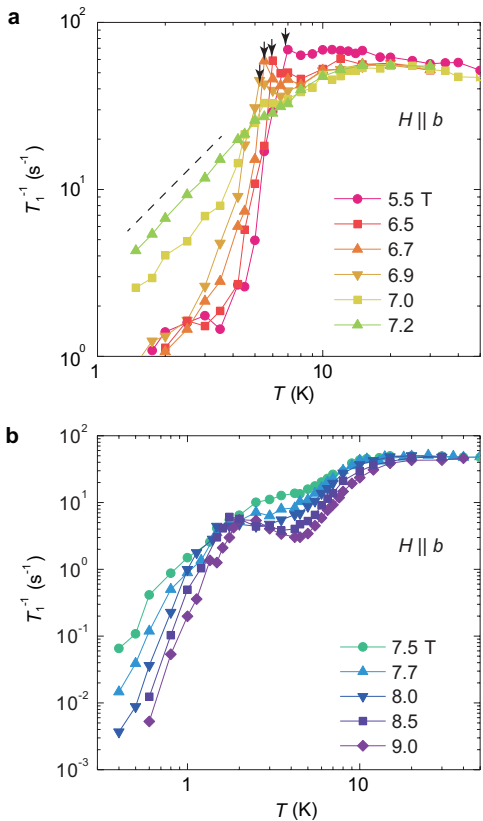}
	\end{center}
	\caption{\label{Fig3} 
	{\bf Nuclear spin-lattice relaxation rate $T_1^{-1}$ under finely tunned magnetic field}. Temperature dependence of $T_1^{-1}$ ({\bf a}) in a low field range below 7.2 T and ({\bf b}) a high field range above 7.5 T. A sharp peak is observed at the magnetic ordering temperature $T_{\rm N}$ marked by arrows, below which $T_1^{-1}$ decreases exponentially. Around $H_{c^\prime}$, $T_1^{-1}$ obeys the $T^{3/2}$ dependence represented as the dotted line. 
	}
	\end{figure}

	The in-plane magnetic field dependence of $T_1^{-1}$ is investigated in details around the critical field $H_c$, as shown in Fig. 3. In a low field range (Fig. 3{\bf a}), $T_1^{-1}$ exhibits a broad maximum around 15--20 K, which is shifted to higher temperatures with increasing $H$. The $T$ dependence is distinct from conventional antiferromagnets showing a power-law evolution toward $T_{\rm N}$ but similar to the behaviour expected for the Kitaev model where the spin gap opens in the ground state\cite{Kitaev}. Above $T_{\rm N}$, we evaluate the spin gap by fitting $T_1^{-1}$ into an extended exponential function $T_1^{-1} \propto  T^{-n}$e$^{(-\Delta_{\rm H}/T)}$ ($n = 1$, Fig. 4{\bf a}) for $10 < T < 30 $ K (Ref. \onlinecite{Jansa}). The obtained gap $\Delta_{\rm H}$ obeys a relation $\Delta_{\rm H} \sim \Delta_0 + bH^\delta$ with the power $\delta = 3.0 \pm 0.3$ for $H>5.5 $ T (Fig. 4{\bf b}) in agreement with the Kitaev's prediction\cite{Kitaev} showing $H^3$ dependence of the MF gap. An extrapolation to zero field yields a constant value $\Delta_0 = 7 \pm 1$ K, close to the flux gap $0.065J_K \sim 10$ K (Ref. \onlinecite{Kitaev}) and $\Delta_0$ estimated from the angular dependence of $T_1^{-1}$ (Ref. \onlinecite{Jansa}). 

In the magnetically ordered state, we observed a steep $T_1^{-1}$ drop due to gapped spin-wave excitations in presence of the strong magnetic anisotropy. We evaluated the magnon gap $\Delta_{\rm m}$ based on the three-magon process with $T_1^{-1} = AT^2$e$^{(-\Delta_m/T)}$ (Fig. S4)\cite{Narath}. The obtained $\Delta_{\rm m} = 60$ K at zero field is suppressed toward $H_c^\prime = 7.10 \pm 0.02$ T (Fig. 4{\bf b}) following $\Delta_{\rm m} \propto (1 - H/H_c^\prime)^{z\nu}$ ($z\nu = 0.36 \pm 0.03$). Here $H_c^\prime$ is slightly lower than $H_c = 7.40$ T for $T_{\rm N}$ (Fig. 4{\bf c}, Fig. S7), as obtained from the $T$ and $H$ dependence of $T_1^{-1}$. The vanishing magnon gap below $H_c$ indicates emergence of a gapless magnetic phase with strong spin fluctuations, where $T_1^{-1}$ obeys a power law $T_1^{-1} \sim T^n$ ($n \sim $ 1.5). As $\Delta_{\rm H}$ overcomes $\Delta_{\rm m}$ around 7 T, the spin liquid state persists down to low temperatures. At $H = 7.5$ T just above $H_c$, we still find power-law $T$ dependence of $T_1^{-1}$ ($n \sim 2$) below 4 K, implying a linear Dirac dispersion with a tiny gap down to low energies. 

For $H > H_c$, $T_1^{-1}$ continues to be suppressed, and a prominent peak around $1-2$ K is shifted to higher temperatures (Fig. 3{\bf b}). Thus, a lower-lying gap also increases with $H$. We can evaluate the spin gap $\Delta_{\rm L}$ using the same equation as for $\Delta_{\rm H}$ below 4 K (Fig, 4{\bf a}). As shown in Fig. 4{\bf b}, $\Delta_{\rm L}$ is suppressed toward $H_c$. The origin of $\Delta_{\rm L}$ is unlikely attributed to a forced ferromagnetic state, because there is no indication of the saturated magnetization in the $K$ measurements below 9 T (Fig. S7). The peak structure of $T_1^{-1}$ around 2 K indicates a sizable spectral weight above $\Delta_{\rm L}$ ($\sim 0.2$--0.5 meV). The energy scale of $\Delta_{\rm L}$ is lower than the spin gap at $q = 0$ as observed in the electron spin resonance\cite{Sears, Ponomaryov, Winter}.

	\begin{figure*}
	\begin{center}
	\includegraphics[scale=0.7]{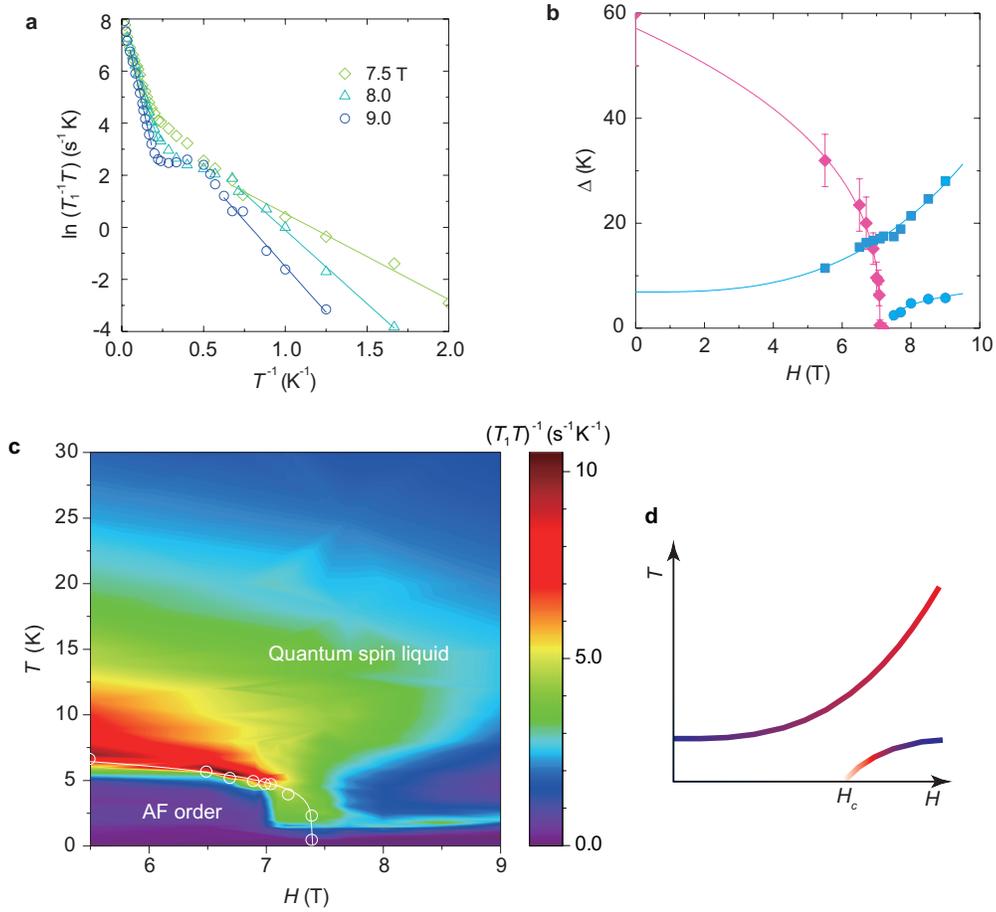}
	\end{center}
	\caption{\label{Fig4} 
	{\bf Criticality of low-lying spin excitations.} {\bf a}, Inverse temperature dependence of ln$(TT_1^{-1})$. The slopes yield the spin gap, $\Delta_{\rm H}$ and $\Delta_{\rm L}$, for temperature ranges, $7-20$ K and $0.5-4$ K, respectively. {\bf b}, Magnetic field dependence of the magnon gap $\Delta_{\rm m}$ (red diamond), $\Delta_{\rm H}$ (blue square), and $\Delta_{\rm L}$ (blue circle). {\bf c}, Contour plot of $(T_1T)^{-1}$ as a function of magnetic field $H$ and temperature $T$. Circles denote $T_{\rm N}$ defined by the $T_1^{-1}$ peak against $H$ and $T$. A solid curve is a fitting result with $T_{\rm N} \sim (1 - H/H_c)^\nu$, yielding $H_c = 7.40 \pm 0.03$ T and the critical exponent $\nu = 0.22 \pm 0.02$. {\bf d}, Schematic energy diagram of fractionalized gapped excitations. The dominant Majorana and gauge flux contributions are colored with red and blue, respectively. 
	}
	\end{figure*}

The quantum criticality of low-lying excitations is visualized by plotting $(T_1T)^{-1}$ as a function of $T$ and $H$ in the contour plot (Fig. 4{\bf c}, Fig. S8). In quantum critical magnets, $(T_1T)^{-1}$ or the dynamical spin susceptibility $\chi({\bf q}, \omega)$ would be divergently enhanced for $T \to 0$ at $H_c$, as the spin correlation length $\xi$ grows with a power law, $\xi \sim T^{-\nu}$ ($\nu = 1$ for two-dimensional Ising magnets). Since $(T_1T)^{-1}$ relates to $\xi$ with $(T_1T)^{-1} \propto \xi^{z-\eta}$ based on the dynamical scaling hypothesis\cite{Hohenberg, Sachdev}, the absence of the critical enhancement in $(T_1T)^{-1}$ shows that the spin correlation remains finite due to the development of $\Delta_{\rm H}$ over the {\bf q} space. 

In the Kitaev model, the spin excitation is expressed with a composite of the gapless MF and the gapped fluxes, resulting in the nonzero spin gap from the ground state to the excited states at zero field. When the magnetic field is introduced, the MF excitation is also gapped, and furthermore the two excitations are not defined independently and entangled with each other. Thus, it is natural to observe two spin gaps in the presence of the magnetic field. In addition, an anti-crossing between the two excitations is expected to occur at some field, as the perturbation theory suggests less $H$ dependence of the flux gap and a rapid increase of the MF gap proportional to $H^3$ (Ref. \onlinecite{Kitaev}). Interestingly, as shown in Fig. 4{\bf b}, $\Delta_{\rm H}$, which originally reflects the flux character at $H \to 0$, has a concave $H$ dependence, while $\Delta_{\rm L}$ is convex. This appears to be consistent with the anticipated anti-crossing behaviour. In this scenario, $\Delta_{\rm L}$ is ascribed to the excitation that originally reflects the MF character but gradually acquires the flux character while increasing $H$ (see Fig. 4{\bf d}). Thus, our observation of the two-step gap opening would be a direct evidence of the two-types of entangled excitations, although it is obviously necessary to clarify the effect of non-Kitaev interactions for complete understanding, including the critical behavior at $H_c$.  

Our observations of the local spin susceptibility and low-energy spin dynamics demonstrate the two types of spin excitations in the field-driven quantum spin liquid state. It strongly suggests the emergent entanglement between the fractionlized MF and $Z_2$ fluxes under the magnetic field. The recombination process and the quantum criticality are the intriguing future issue requiring further extensive investigations beyond the analytically soluble Kitaev model. 


 

\noindent{\bf Crystal growth and characterization} 
Single crystals of $\alpha$--RuCl$_3$ were grown by a sublimation method. The starting material of anhydrous ruthenium trichloride was dried in vacuum at 250 $^\circ$C for 24 hours. The powder sealed in a quartz tube was heated up to 1100 $^\circ$C and slowly cooled down to 600 $^\circ$C under the spacial thermal gradient. The plate crystals were obtained in the cooler side of the tube. Since the crystal might cause metastable stacking faults by applying stress, we took special care for handing the crystal. The quality of the single crystal was evaluated by measuring x-ray diffraction, NMR, NQR, and magnetic susceptibility. The magnetization was measured with a superconducting quantum interference device (SQUID) magnotometer (MPMS-XS, Quantum Design ltd.) at 1 T. The magnetic transition was observed only at $T_{\rm N}$ = 6.5 K in magnetic susceptibility and NQR measurements, signifying a single domain crystal without a stacking fault. 

\noindent{\bf Nuclear quadrupole and magnetic resonance}  
We performed $^{35}$Cl and $^{37}$Cl NQR measurements on a single crystal at zero field down to 1.4 K. NMR was measured at constant magnetic fields up to 9.0 T in a temperature range of 0.4--300 K by utilizing a $^3$He cryostat below 1.5 K. The NQR and NMR spectra were obtained from Fourier transformation of temporal spin-echo evolution after $\pi/2$ and $\pi$ pulses with an interval time $\tau = 40$ $\mu$s. The nuclear spin-lattice relaxation rate $T_1^{-1}$ was obtained from a nuclear magnetization recovery fitted to an exponential function $	\frac{M(\infty )-M(t)}{M(\infty )} = 0.1{\rm e}^{-t/T_1} + 0.9{\rm e}^{-6t/T_1}$ for the central NMR line or fitted to a stretched exponential function below 10 K. The spin-echo decay rate $T_2^{-1}$ was obtained from the $2\tau$ dependence of the integrated spin-echo intensity. Beyond the second-order approximation of electric quadrupole contribution to the spectral shift, the $^{35}$Cl Knight shift was evaluated by analyzing the central line position with the exact diagonalization of the nuclear spin Hamiltonian $\mathcal H = {\bf S}\cdot {\sf A}\cdot {\bf I} + {\bf I}\cdot {\sf q} \cdot {\bf I}$ for the electron spin {\bf S}, the hyperfine coupling tensor ${\sf A}$, and the electric quadrupole coupling tensor ${\sf q}$ by considering the temperature dependence of the NQR frequency (Fig. S1). Further details of the spectral assignment were described in Supplementary Information.





\begin{thebibliography}{99}
\bibitem{Kitaev}
Kitaev, A. 
Anyons in an exactly solved model and beyond. 
{\it Ann. Phys. (Amsterdam)} {\bf 321}, 2--111 (2006). 

\bibitem{Plumb}
Plumb, K. W. {\it et al.} 
 $\ensuremath{\alpha}\ensuremath{-}{\mathrm{RuCl}}_{3}$: A spin-orbit assisted Mott insulator on a honeycomb lattice. 
{\it Phys. Rev. B} {\bf 90}, 041112 (2014).

\bibitem{Kubota}
Kubota, Y. {\it et al.} 
Successive magnetic phase transitions in $\ensuremath{\alpha}\ensuremath{-}{\mathrm{RuCl}}_{3}$: XY-like frustrated magnet on the honeycomb lattice. 
{\it Phys. Rev. B} {\bf 91}, 094422 (2015).

\bibitem{Sandilands}
Sandilands, L. J., Tian, Y., Plumb, K. W., $\&$ Kim, Y.-J. 
Scattering continuum and possible fractionalized excitations in $\ensuremath{\alpha}\text{\ensuremath{-}}{\mathrm{RuCl}}_{3}$. 
{\it Phys. Rev. Lett.} {\bf 114}, 147201 (2015).

\bibitem{Chaloupka}
Chaloupka, J. $\&$ Khaliullin, G. 
Magnetic anisotropy in the Kitaev model systems ${\mathrm{Na}}_{2}{\mathrm{IrO}}_{3}$ and ${\mathrm{RuCl}}_{3}$.
{\it Phys. Rev. B} {\bf 94}, 064435 (2016). 

\bibitem{Banerjee}
Banerjee, A.  {\it et al.} 
Proximate Kitaev quantum spin liquid behaviour in a honeycomb magnet. 
{\it Nat. Mater.} {\bf 15}, 733--740 (2016).

\bibitem{Banerjee2}
Banerjee, A. {\it et al.} 
Neutron scattering in the proximate quantum spin liquid $\ensuremath{\alpha}\ensuremath{-}{\mathrm{RuCl}}_{3}$, 
{\it Science} {\bf 356}, 1055--1059 (2017). 

\bibitem{Do}
Do, S.-H.  {\it et al.} 
Majorana fermions in the Kitaev quantum spin system $\alpha$-RuCl$_3$, 
{\it Nat. Phys.} {\bf 13}, 1079--1084 (2017).

\bibitem{Wolter}
Wolter, A. U. B. {\it et al.} 
Field-induced quantum criticality in the Kitaev system $\ensuremath{\alpha}\ensuremath{-}{\mathrm{RuCl}}_{3}$. 
{\it Phys. Rev. B} {\bf 96}, 041405 (2017).

\bibitem{Baek}
Baek, S.-H. {\it et al.} 
Evidence for a field-induced quantum spin liquid in $\ensuremath{\alpha}$-${\mathrm{RuCl}}_{3}$. 
{\it Phys. Rev. Lett.} {\bf 119}, 037201 (2017).

\bibitem{Sears}
Sears, J. A. {\it et al.} 
Phase diagram of $\ensuremath{\alpha}\ensuremath{-}{\text{RuCl}}_{3}$ in an in-plane magnetic field. 
{\it Phys. Rev. B} {\bf 95}, 180411(R) (2017).

\bibitem{Ponomaryov}
Ponomaryov, A. N. {\it et al.} 
Unconventional spin dynamics in the honeycomb-lattice material $\alpha$-RuCl3: High-field electron spin resonance studies, 
{\it Phys. Rev. B} {\bf 96}, 241107(R) (2017). 

\bibitem{Winter}
Winter, S. M. {\it et al.} 
Probing $\alpha$-RuCl$_3$ beyond magnetic order: effects of temperature and magnetic field, 
{\it Phys. Rev. Lett.} {\bf 120}, 077203 (2018).

\bibitem{Knolle}
Knolle, J. Kovrizhin, D. L. Chalker, J. T. $\&$ Moessner, R. 
Dynamics of a two-dimensional quantum spin liquid: signatures of emergent Majorana fermions and fluxes. 
{\it Phys. Rev. Lett.} {\bf 112}, 207203 (2014).

\bibitem{Song}
Song, X. Y., You, Y. Z., $\&$ Balents, L. 
Low-energy spin dynamics of the honeycomb spin liquid beyond the Kitaev limit,
{\it Phys. Rev. Lett.} {\bf 117}, 037209 (2016). 

\bibitem{Yoshitake}
Yoshitake, J., Nasu, J. $\&$ Motome, Y. 
Fractional spin fluctuations as a precursor of quantum spin liquids: Majorana dynamical mean-field study for the Kitaev model. 
{\it Phys. Rev. Lett.} {\bf 117}, 157203 (2016).

\bibitem{Gohlke}
Gohlke, M., Verresen, R., Moessner, R., $\&$ Pollmann, F.  
Dynamics of the Kitaev-Heisenberg model, 
{\it Phys. Rev. Lett.} {\bf 119}, 157203 (2017). 

\bibitem{Laughlin}
Laughlin, R. B. 
Anomalous quantum Hall effect: an incompressible quantum fluid with fractionally charged excitations. 
{\it Phys. Rev. Lett.} {\bf 50}, 1395--1398 (1983). 

\bibitem{Jackeli}
Jackeli G. $\&$ Khaliullin, G. 
Mott insulators in the strong spin-orbit coupling limit: from Heisenberg to a quantum compass and Kitaev models. 
{\it Phys. Rev. Lett.} {\bf 102}, 017205 (2009).


\bibitem{Nasu}
Nasu, J. Udagawa, M. $\&$ Motome, Y. 
Thermal fractionalization of quantum spins in a Kitaev model: Temperature-linear specific heat and coherent transport of Majorana fermions. 
{\it Phys. Rev. B} {\bf 92}, 115122 (2015).








\bibitem{Yoshitake2}
Yoshitake, J., Nasu, J. $\&$ Motome, Y. 
Majorana dynamical mean-field study of spin dynamics at finite temperatures in the honeycomb Kitaev model. 
{\it Phys. Rev. B} {\bf 96}, 024438 (2017). 

\bibitem{Yoshitake3}
Yoshitake, J., Nasu, J. $\&$ Motome, Y. 
Temperature evolution of spin dynamics in two- and three-dimensional Kitaev models: Influence of fluctuating $Z_2$flux.  
{\it Phys. Rev. B} {\bf 96}, 064433 (2017). 



\bibitem{Wang}
Wang, Z. 
Magnetic excitations and continuum of a possibly field-induced quantum spin liquid in $\ensuremath{\alpha}\text{\ensuremath{-}}{\mathrm{RuCl}}_{3}$,
{\it Phys. Rev. Lett.} {\bf 119}, 227202 (2017). 



\bibitem{Wellm}
Wellm, C. {\it et al.} 
Signatures of low-energy fractionalized excitations in $\ensuremath{\alpha}\ensuremath{-}{\mathrm{RuCl}}_{3}$ from field-dependent microwave absorption, 
arXiv:1710.00670v. 

\bibitem{Zheng}
Zheng, J. 
Gapless spin excitations in the field-induced quantum spin liquid phase of $\alpha$-RuCl$_3$, 
{\it Phys. Rev. Lett.} {\bf 119}, 227208 (2017). 

\bibitem{Jansa}
Jansa, N. {\it et al.} 
Observation of gapped anyons in the Kitaev honeycomb magnet under a magnetic field. 
{\it Nat. Phys.} {\bf 14}, 786 -- 790 (2018). 


\bibitem{Morosin}
Morosin, B. $\&$ Narath, A. 
X-ray diffraction and nuclear quadrupole resonance studies of chromium trichloride. 
{\it J. Chem. Phys.} {\bf 40}, 1958 (1964).

\bibitem{Moriya}
Moriya, T. 
Nuclear magnetic relaxation in antiferromagnetics. 
{\it Prog. Theo. Phys. }{\bf 16}, 23--44 (1956).

\bibitem{Hohenberg}
	Hohenberg, P. C. $\&$ Halperin, P. I. 
	Theory of dynamic critical phenomena, 
	{\it Rev. Mod. Phys. } {\bf 49}, 435--479 (1977). 

\bibitem{Dutta}
Dutta, A. {\it et al.} 
{\it Quantum Phase Transitions in Transverse Field Spin Models: From Statistical Physics to Quantum Information}
(Cambridge University Press, Cambridge, 2015). 

\bibitem{Narath}
Narath A. $\&$ Fromhold, A. T. 
Nuclear spin-lattice relaxation of $^{53}$Cr in the ordered magnetic insulator CrCl$_3$, 
{\it Phys. Rev. Lett} {\bf 17}, 354 (1966).

\bibitem{Sachdev}
	Sachdev, S. {\it Quantum Phase Transitions}, 2nd. Ed. (Cambridge Univ. Press. London, 2011). 


\end{thebibliography}
\end{document}